\documentclass[review]{elsarticle}
\usepackage{lineno,hyperref}
\modulolinenumbers[5]    
\usepackage{mathtools}
\usepackage{enumerate}
\usepackage{amsmath}
\usepackage{siunitx}
\usepackage{inputenc}
\usepackage{amsmath,amssymb,latexsym,epstopdf,array,algorithm,algpseudocode}
\usepackage{array}
\usepackage{caption}
\usepackage{float}
\usepackage{multirow,booktabs,colortbl,tabularx}
\usepackage{tikz}
\usetikzlibrary{arrows,positioning,shapes}
\newcommand*{\Comb}[2]{{}^{#1}C_{#2}}%

\begin{document}
	
	\begin{frontmatter}
		
		\long\def\symbolfootnote[#1]#2{\begingroup%
			\def\thefootnote{\fnsymbol{footnote}}\footnote[#1]{#2}\endgroup}

\title{A fingerprint based crypto-biometric system for secure communication}
\author[mymainaddress]{Rudresh Dwivedi\corref{mycorrespondingauthor}}
\cortext[mycorrespondingauthor]{Corresponding author}
\ead{phd1301201006@iiti.ac.in}

\author[mymainaddress]{Somnath Dey}
\ead{somnathd@iiti.ac.in}
\author[mymainaddress]{Mukul Anand Sharma}

\author[mymainaddress]{Apurv Goel}

\address[mymainaddress]{Discipline of Computer Science \& Engineering,  \\
	Indian Institute of Technology Indore, Indore, India.}

	\begin{abstract}
	To ensure the secure transmission of data, cryptography is treated as the most effective solution. Cryptographic key is an important entity in this procedure. In general, randomly generated cryptographic key (of 256 bits) is difficult to remember. However, such a key needs to be stored in a protected place or transported through a shared communication line which, in fact, poses another threat to security. As an alternative, researchers advocate the generation of cryptographic key using the biometric traits of both sender and receiver during the sessions of communication, thus avoiding key storing and at the same time without compromising the strength in security. Nevertheless, the biometric-based cryptographic key generation possesses few concerns such as privacy of biometrics, sharing of biometric data between both communicating users (i.e., sender and receiver), and generating revocable key from irrevocable biometric. This work addresses the above-mentioned concerns. 
	
	In this work, a framework for secure communication between two users using fingerprint based crypto-biometric system has been proposed. For this, Diffie-Hellman (DH) algorithm is used to generate public keys from private keys of both sender and receiver which are shared and further used to produce a symmetric cryptographic key at both ends. In this approach, revocable key for symmetric cryptography is generated from irrevocable fingerprint. The biometric data is neither stored nor shared which ensures the security of biometric data, and perfect forward secrecy is achieved using session keys. This work also ensures the long-term security of messages communicated between two users. Based on the experimental evaluation over four datasets of FVC2002 and NIST special database, the proposed framework is privacy-preserving and could be utilized onto real access control systems. 
\end{abstract}

\begin{keyword}
Biometric security \sep Diversity \sep Fingerprint \sep Minutiae \sep Revocability \sep Template security
\end{keyword}

\end{frontmatter}

%\linenumbers

\section{Introduction}
\label{sec1}
\subsection{Background}
The identity of a user is lost if user's original biometric information is compromised. The biometric systems which integrate biometrics with cryptography are called crypto-biometric systems \cite{bioc1}. For better security of a cryptographic system, keys used for encryption and decryption must be long enough to be unbreakable. Knowledge-based (the key is remembered by the user) and possession-based (key stored in smart card etc.) authentication systems are not secure due to the fact that long keys cause user inconvenience to remember and smart cards can be stolen or misplaced. Moreover, storing long keys on a system is costly and not secure. Biometrics-based authentication systems can alleviate the limitations of above-mentioned systems \cite{bioc}. A user's biometric is integrated with cryptography using either key-generation techniques in which cryptographic key is generated from one's biometric or key-binding techniques in which cryptographic key is fused to the original biometric data \cite{survey}. When user A wants to send a message to user B, A first encrypts the message using a key $K$ and then sends this encrypted message to B. B can decrypt this message using key $K$ only. For this, either the key $K$ or some information to generate the same key $K$ at both ends (A and B) must be shared between two communicating users. In both cases, sharing of some information is required. Therefore, there is need of securely sharing information over the non-secure communication channel.

\subsection{Existing approaches}
In biometric cryptosystems (BCs), biometric data is combined using cryptographic keys to provide security and privacy in user's authentication. As described in Section 1, the cryptographic keys are derived from biometric data of end users using a hash function (one way) or user defined algorithm in key generation schemes whereas key-binding systems transform the biometric information using key. 

Few of the approaches have been proposed to generate cryptographic key from biometric traits \cite{monrose,hao,irisc,chen}. Monrose et al. \cite{monrose} proposed a technique which records a user's voice while speaking a password. Different segments of a password are mapped to a random look-up table to derive the cryptographic key. Hao et al. \cite{hao} proposed a technique which incorporates the dynamic information like velocity, pressure, altitude, and azimuth. Feature coding was used to quantize each feature into bits which were concatenated to form a cryptographic key. Chen et al. \cite{chen} utilized radon transform onto 3D face data to produce 1-D bit string. Further, keys of suitable length for 128-bit Advanced Encryption Standard (AES) are derived. Rathgeb et al. \cite{irisc} proposed a technique which derives iriscodes using the method implemented by Masek \cite{libor}. Next, the most stable bits within iris codes are selected and their positions are utilized to construct biometric keys. 

In key-binding schemes, Soutar et al. \cite{mytec} proposed a technique which links the biometric feature string with an $N$-bit cryptographic key. During linking, redundancy is added by applying a repetitive code structure. Next, the hash of cryptographic key is stored along with template for secure authentication. Juels et al. \cite{fucom} introduced a fuzzy commitment scheme which binds a codeword to the witness (biometric data). The hash values are stored as the commitment for authentication. Hao et al. \cite{bioc1} applied the fuzzy commitment scheme onto 2048-bit iris-codes. Next, Hadamard and Reed-Solomon error correction codes are utilized to correct bit errors. The fuzzy vault scheme \cite{vault} is the most popular technique for the key-binding schemes. The main idea is to utilize the biometric information to lock a secret key. Clancy et al. \cite{clancy} applied the fuzzy vault scheme onto a set of minutiae points of a fingerprint. The minutiae positions are mapped to a polynomial and chaff points are added to construct a random vault. Reed-Solomon codes are applied to reconstruct the polynomial secure authentication. Kanade et al. \cite{threef} proposed a three-factor key regeneration for iris-based authentication system. A user-specific shufﬂing key derived using a password is used to randomize the iris code. Iris code shufﬂing reduces the error structure in the iris code since the errors gets spread out.  Further, Hadamard codes are used for correcting remaining bit-errors. In another work, Kanade et al. \cite{session} proposed cancelable biometric system with fuzzy commitment based key regeneration scheme. First, a key is randomly generated and then encoded into a pseudo code using Error Correcting Codes (ECC). A cancelable transformation is applied on the reference biometric data of the user. This transformed data is then XORed with the pseudo code to obtain a locked code. 

In present era, researchers are working onto key management for biometric based authentication. Very little work has been proposed about a framework for secure communication on a network using crypto-biometric system. Barman et al. \cite{barman} proposed a system in which both sender and receiver exchange their cancelable biometrics using key-based steganography. Kanade et al. \cite{kana} proposed a crypto-biometric system for establishing secure communication session between two clients. Their method involves CARA (Central Authority for Registration and Authentication) with which the clients are registered. Barman et al. \cite{barman2} proposed a key exchange protocol to integrate the biometric data of two communicating users to bind a secret key (i.e., the session key) that is used for secure message communication. First, the fingerprint data is converted into a binary string to generate a cancelable template. These bit-strings are used to generate mutual lockers and personalized lockers. Cryptographic keys are protected and exchanged using these lockers. Panchal et al. \cite{barman3} proposed a technique in which a unique code is derived from original fingerprint features using the convolution coding principle. Next, the unique code is used to generate a cryptographic key for encryption and decryption of the user's document. 

\subsection{Motivation and contributions}
In key generation schemes, the following issues may arise. They are (i) deformations in the biometric data may derive an erroneous key, (ii) generation of a cryptographic key may require the transmission of biometric data over a network and (iii) there is a need of revocable keys since biometric data is irrevocable and irreplaceable. In key-binding based schemes, errors in the biometric data results to derive erroneous helper data affecting the overall performance of authentication system. The crypto-biometric system providing secure communication onto a network also have some limitations. Storing of biometric templates is one of such issues which should be avoided. Further, one time password (OTP) based communication requires that a user has to remember it for the whole session. Also, compromise may results into privacy elusion.

A crypto-biometric system for secure communication among different users require (i) the generation of unique cryptographic keys from both sender and receiver, (ii) secure transmission of keys among users and (iii) should be secure from all such possible attacks. Moreover, it should also provide privacy to biometrics of users along with generating revocable and non-invertible cryptographic key from biometric data of users.
This work aims to address above mentioned concerns. In this work, a complete framework for secure communication among users on a network using crypto-biometric system has been proposed to provide perfect forward secrecy. \textbf{The sedulous contribution of our method is that we do not store the cryptographic key anywhere. Also, there is no need to store the original biometric template of a user to generate the key. Therefore, there is no overhead of maintaining the cryptographic in our approach.} The contributions of our work are described as follows:
\begin{enumerate}
	\item In this approach, public key cryptography has been used to generate symmetric cryptographic key from fingerprints of users. For this, the DH algorithm of public key cryptography has been used to generate symmetric cryptographic keys. 
	\item The proposed system ensures the avoidance of biometric template storage or cryptographic key, either in a central database or a smart-card. 
	\item The proposed system fulfills the requirement of generating a revocable and non-invertible biometric template to provide secure communication between sender and receiver. 
	\item Use of a central authority and public key cryptography algorithms such as DH and RSA provide security against various attacks including the Man-in-the-middle (MiM) attack.
\end{enumerate}
The remainder of this paper is organized as follows. A brief review of existing related work is described in section 2. Section 3 describes the proposed approach in detail. Experimental results and security analysis of this method are presented in section 3 and 4, respectively. Finally, the paper is concluded in section 5 with a glimpse on future direction.

\section{Proposed methodology}
The proposed work initially extracts pair-minutiae features from the sender and receiver. A binary string is obtained after quantization and binning. Next, a random key based permutation is applied on feature bit string to obtain a permuted binary string. This binary string is hashed using SHA256 to generate a 256-bit private key which is used as an input to DH algorithm along with two predefined parameters to generate public keys of sender and receiver. These public keys are then shared between sender and receiver. DH algorithm uses user's own private key and other user's public key to generate a symmetric key at both users end. This key is termed the intermediate key which is further hashed to generate final cryptographic key. This key is then used for encryption and decryption of information to be shared between sender and receiver.

This system also involves authentication of users before starting communication among them. For this, a central authority (CA) for enrollment and verification of users has been proposed. At the time of registration, a user generates an RSA public-private key pair and shares this public key with CA along with some identification. CA registers the user with all this information and provides a signed certificate to the user. This certificate is used by users to verify each other before setting up the connection as described in Fig. 1.

\begin{figure}[!htbp]
	\centering
	\includegraphics[width=3.55in, height=2.4in]{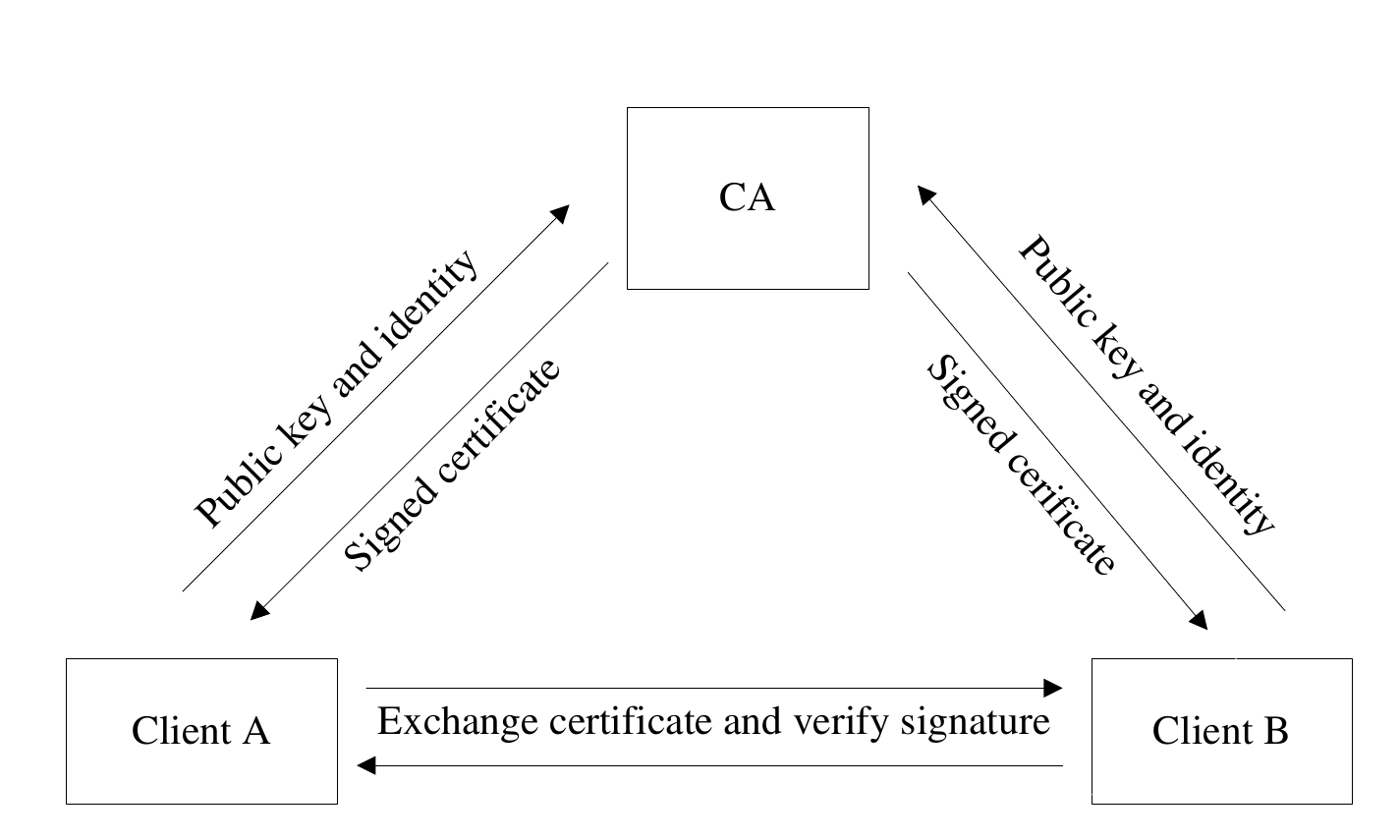}
	\caption{Enrollment and Authentication using CA}
\end{figure}

The above-mentioned steps of the proposed framework are stated in detail in the following subsections. Fig. 2 gives the detailed diagram of the proposed framework.

\begin{figure}[!htbp]
	\centering
	\includegraphics[width=\textwidth, height=5.5in]{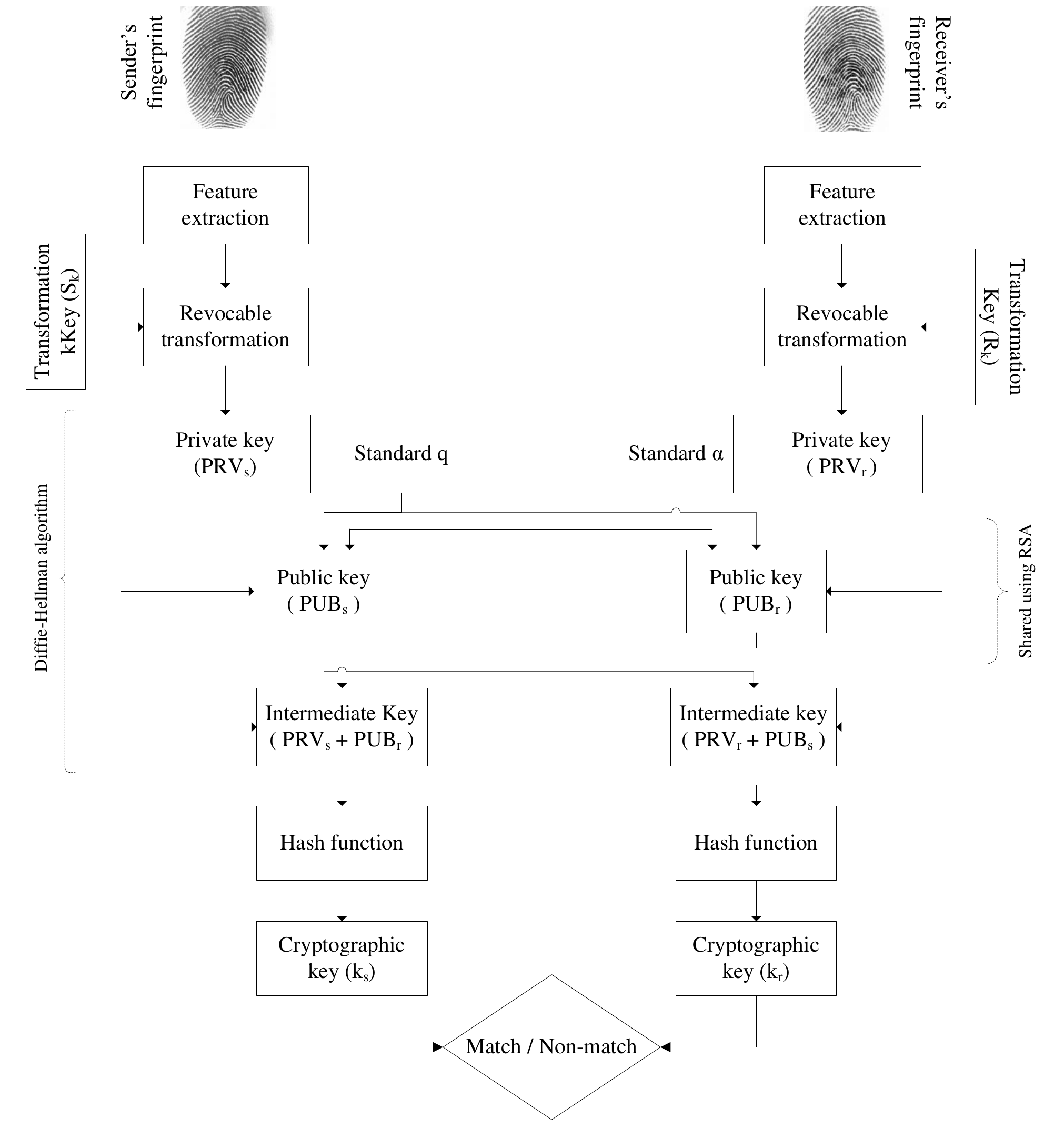}
	\caption{Proposed Crypto-biometric system framework}
\end{figure}

\subsection{Feature extraction}
The performance of the fingerprint based verification system may degrade due to by rotation, translation and scaling deformations caused at the time of image acquisition. Hence, there is a need to evaluate translation and rotation-invariant features from the fingerprint image. For this purpose, we utilize the pair-minutiae feature extraction technique which was originally proposed by Jin et al. in \cite{ks2}. For better understanding, we briefly describe this procedure. Let, set of minutiae points extracted from fingerprint image are denoted as:

\begin{equation}\label{eq:e1}
Ms = \left \{ Ms_{k}\left ( x_{k},y_{k},\theta_{k} \right ) \right \}^{n}_{k=1}
\end{equation}
where, $n$ is number of minutiae points. ($x_{k}, y_{k}, \theta_{k}$) are x, y coordinates and orientation of $k^{th}$ minutiae, respectively. A pair minutiae vector $Vp_{ij}$ can be formed by pairing up two minutiae $Ms_{i}$ and $Ms_{j}$ from set $Ms$. There will be $\frac{n(n-1)}{2}$ pairs constituting the set $Vp$ which can be expressed as:

\begin{equation}\label{eq:e2}
Vp = \{Vp_{ij} : {1}\leq i,j \leq n \ \text{and} \ i\neq j\}
\end{equation}
where, each $Vp_{ij}$ is triplet of distance and relative angles of minutiae pair \textit{($Ms_{i},Ms_{j}$)}, assuming the reference direction of line segment connecting minutiae pair is from $Ms_{i}$ to $Ms_{j}$. Hence, $Vp_{ij}$ is defined as:

\begin{equation}\label{eq:e3}
Vp_{ij} = \{L,\alpha_{i},\beta_{j}\}
\end{equation}
where, \textit{L} is the distance between minutiae pairs $Ms_{i}$ and $Ms_{j}$. $\alpha_{i}$ is the angle between reference direction of line segment joining $Ms_{i}$ and $Ms_{j}$ with the orientation of $Ms_{i}$ in the counter-clockwise direction; $\beta_{j}$ is defined analogously. Figure 3 illustrates this triplet formation.

\begin{figure}[!htbp]
	\centering
	\includegraphics[width=3in, height=1.5in]{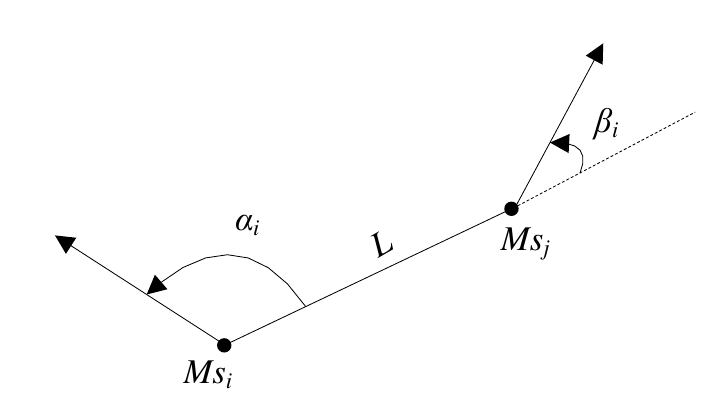}
	\caption{Pair-minutiae feature extraction}
\end{figure}

To determine $Vp_{ij}$, the following two quantities \textit{X} and \textit{Y} are calculated first:

\begin{center}
	$X = (x_{j}-x_{i})cos\theta_{i} + (y_{j}-y_{i})sin\theta_{i}$
\end{center}
\begin{center}
	$Y = (x_{j}-x_{i})sin\theta_{i} - (y_{j}-y_{i})cos\theta_{i}$
\end{center}

Next, the triplet contained in $Vp_{ij}$ = ($L$, $\alpha_{i}$, $\beta_{j}$) is obtained as follows:

\begin{equation}\label{eq:e6}
L = \sqrt{X^{2} + Y^{2}}
\end{equation}

\begin{equation}
\alpha_{i} = \text{arctan} \left (\frac{Y}{X}  \right )
\end{equation}

\begin{equation}\label{eq:e8}
\beta_{j} = \alpha_{i} + \theta_{j} - \theta_{i}
\end{equation}

After the evaluation of ($L$, $\alpha_{i}$, $\beta_{j}$), quantization is applied on each $Vp_{ij}$ in $Vp$. \textit{($L$, $\alpha_{i}$, $\beta_{j}$)} are represented in binary notation by choosing a quantization step size. Suppose $n_{l}$, $n_{\alpha}$ and $n_{\beta}$ are number of bits required to represent $L$, $\alpha$ and $\beta$ in binary notation, respectively. Then the total number of bits to represent each $Vp_{ij}$ in $Vp$ are represented as:

\begin{equation}
n_{p} = n_{l} + n_{\alpha} + n_{\beta}
\end{equation}

Thus, for each pair-minutiae vector $Vp_{ij}$ in $Vp$, a bit-string $Vp_{ij}^{(b)}$ of $n_{p}$ bits is derived. The set $Vp^{(b)}$ represents the set of $Vp_{ij}^{(b)}$ as follows: 

\begin{equation}\label{eq:e10}
Vp^{(b)} = \{Vp_{ij}^{(b)} : {1}\leq i,j \leq {n} \ \text{and} \ {i}\neq j\}
\end{equation}

Empirical evaluations find that $n_{p}$=15 (i.e. 5 bits for each $L$, $\alpha$ and $\beta$ provides the optimal equal error rate (EER) and the maximum entropy [Please see Section 3.6]. Further, binning is applied on binarized pair-minutiae vector set. Since there are $2^n$ possible combinations of n bits, binning starts from 00 $\cdots$ 0 to 11 $\cdots$ 1. For each \textit{$Vp_{ij}^{(b)}$} in \textit{$Vp^{(b)}$},we index a bin by 1 if \textit{$Vp_{ij}$} falls in it. The bins indexed at least once are assigned 1 and all other bins are assigned 0. At the end of this procedure, a binary string {\textit{h$_{k}$}} of length $2^{n_{p}}$ is obtained in which 1's correspond to the unique occurrence of those \textit{$Vp_{ij}^{(b)}$}. In this work, this binary string $h_{k}$ is considered as the feature vector.

\subsection{Authentication system}
In crypto-biometric systems, transformation either binds or derive a cryptographic key providing revocability and non-invertibility to the original biometric. The user authentication is performed using this cryptographic key which can easily go in hands of adversaries. However, the proposed framework does not share or store biometric data. DH algorithm is used for generating symmetric keys at both sender and receiver ends. We describe the procedure in the following subsections.

\subsubsection{Revocable transformation}
To provide revocability to the bit string, random permutation is applied on the bit string based on the user-specific key. This key is termed as transformation key for the user. The transformation key is used as seed value to generate random numbers equal to length of feature bit string. Bits corresponding to these random numbers are swapped with bits at positions starting from the start and incrementing with each random number. For example, we have a bit string of length 15 bits. The transformation key ($T$) is used as seed to generate random numbers from 1 to 15. Say, first random number generated is 5. We swap bit at $5^{th}$ position with $1^{st}$ position bit. Let, next random number be 11. Now, $2^{nd}$ bit is swapped with $11^{th}$ bit. This process continues till last bit is swapped with bit at position equal to new random number. In this way, a bit string is derived which is the permutation of original bit string. 

The method ensures the generation of a revocable template from feature bit-string of the user's biometric since a user can utilize different values of $T$ to generate a different template from same feature bit string.

\subsubsection{Generation of public key and cryptographic key using DH algorithm}
DH algorithm which is used to enable users to securely exchange a cryptographic key over public channels. We apply DH algorithm using this private key to generate a public key for the users. Further, authentication using CA is carried out using public-private key pair. For convenience, the procedural steps are as follows: 

\begin{algorithm}
	\caption{Diffie-Hellman (DH) Algorithm}\label{alg:DH}
	\begin{algorithmic}[1]
		\State Take a prime number q
		\State Take an integer $\alpha$ such that $\alpha$ $<$ q and $\alpha$ is a primitive root of q
		\State For user A, select a random integer $X_{A}$ $<$ q
		\State Calculate $Y_{A}$ = $\alpha^{X_{A}}$ mod q
		\State For user B, select a random integer $X_{B}$ $<$ q
		\State Calculate $Y_{B}$ = $\alpha^{X_{B}}$ mod q
		\Comment $Y_{A}$ and $X_{A}$ are public and private keys of user A; $Y_{B}$ and $X_{B}$ are public and private keys of user B, respectively
		\State Public keys $Y_{A}$ \& $Y_{B}$ are shared among each other
		\State Calculation of cryptographic key $K_{s}$ by user A as $ K_{s} = (Y_{B})^{X_{A}}$ mod q
		\State Calculation of cryptographic key $K_{r}$ by user B as $ K_{r} = (Y_{A})^{X_{B}}$ mod q		
	\end{algorithmic}
\end{algorithm}

In DH algorithm, it is easy to calculate exponentials modulo a prime while it is very difficult to calculate discrete logarithms. For large primes, the latter task is considered infeasible. As discussed in subsection 2.1, $n$ is considered to be 15 in our approach to get a feature bit string of length $2^{15}$. Same is the length of revocable template which is permuted feature bit string. This large binary string needs to be mapped into a smaller one which can be used as key input for DH algorithm. For this, SHA256 hash has been used here. Permuted binary string is hashed using SHA256 to generate a 256-bit key. This key is termed as private key of the user. This way, private keys of sender ($PRV_{S}$) and receiver ($PRV_{R}$) are generated. 

Further, DH algorithm requires a large prime number q and its primitive root $\alpha$. These parameters are not required to be generated in each session, we can also use fixed value of these parameters over a large number of sessions (Appendix A). In this approach, DH parameters of RFC 3526: 2048-bit MODP group \cite{mme} have been used. With private keys $PRV_{S}$, $PRV_{R}$, q and $\alpha$, step 4 and 5 of DH algorithm are applied to generate public keys $PUB_{S}$ and $PUB_{R}$ of sender and receiver respectively. 

Public keys $PUB_{S}$ and $PUB_{R}$ are then shared between sender and receiver. Once both sender and receiver have each other's public keys, DH algorithm is applied at both ends to generate a secret key using own private key and other's public key, as explained in step 9 and 10 of DH algorithm. This way, both sender and receiver derive a secret key. This key is termed as intermediate key for the communication setup. In our approach, size of this key is 2048 bits. This intermediate key is hashed using SHA256 to generate a 256-bit key which is the final cryptographic key. This key is then used for encryption and decryption of messages between sender and receiver.

\subsubsection{Authentication using CA}
The proposed crypto-biometric system also involves a central authority (CA) with which all users need to be registered. When a new user joins the system, CA requires to enroll it first. Following steps are performed in enrollment phase:
\begin{enumerate}
	\itemsep0em 
	\item User generates its own set of RSA public-private key pair and sends its public key along with its identification to the CA after encrypting it with public key of CA.
	\item CA identifies the user using this information and stores this identification in its database.
	\item CA computes hash of public key and identification of the user and encrypts this hash, public key and identification of the user using its private key. 
	\item This encrypted message is termed as certificate of the user and is sent to the user.
\end{enumerate}

All users enroll with the CA to get their certificates. These certificates are used for verification of other users before setting up a connection with them. Suppose user A wants to communicate with user B. For this purpose, following steps needs to be performed before setting up this connection in verification phase:
\begin{enumerate}
	\itemsep0em 
	\item A sends its certificate to the user B with a request to initiate the communication.
	\item B decrypts A's certificate with CA's public key and computes hash of A's public key and identification. This hash is matched with hash in the certificate to verify that this certificate is indeed signed by the CA.
	\item B then identifies A using its identification and then send its certificate to A.
	\item A does the same steps as B to verify B. Once verified by each other, they can start setting up the communication using proposed approach discussed in above subsections.
\end{enumerate}

\section{Experimental results and analysis}
The proposed framework for secure communication is evaluated based on the two criteria i.e. cryptographic key randomness and performance. In the following subsections, we present the experimental results and performance of the proposed method. We use four performance metrics to evaluate the performance of our method:
\begin{enumerate}
	\itemsep0em 
	\item FAR: The probability of mistakenly accepting an imposter as a genuine user
	\item FRR: The probability of mistakenly rejecting a genuine user as an imposter
	\item GAR: Can be calculated as 1-FRR
	\item EER: The error rate where FAR and FRR hold equality
\end{enumerate}

\subsection{Database}
The proposed method has been evaluated using the four datasets of FVC2002 \cite{fvc} databases (i.e. FVC2002DB1, FVC2002DB2, FVC2002DB3 and FVC2002DB4). Each dataset comprises of 100 subjects with 8 impressions per subject. The performance is evaluated using FVC protocol which states that all possible unique pair of impressions from the same subjects are considered to derive genuine cryptographic keys. As a result, we obtain 2800 (i.e. $\Comb{8}{2}\times 100$) genuine key comparisons. Next, unique pair of impression from different subjects are matched to derive 4950 (i.e. $\Comb{100}{2}$) imposer key comparisons. 

\subsection{Randomness of cryptographic key for genuine pair of subjects} 
To evaluate the randomness cryptographic key for genuine pair of fingerprints, first two fingerprints of each subject from all four FVC2002 datasets DB1, DB2, DB3 and DB4 are  considered as a genuine pair. For each subject, cryptographic key for first two instances of fingerprint are generated. Next, we evaluate number of matching bits between these two keys. Hence, a total of 100 data points are calculated as percentage of matching number of bits out of total number of bits in the generated feature string, for each dataset. A total of $100\times 4$=400 data points are calculated and are illustrated using histograms in Fig 4. It can be observed from the histogram that mean value for matching percentage for a genuine pair of fingerprints is 89.99\% which means that average number of matching bits is 3686 bits out of 4096 bits. The percentage of matching bits is spread between the ranges of 81.37\% to 99.83\% with a standard deviation of 0.043. Therefore, it is evident that a genuine pair achieves sufficient number of matchable bits for the pair of cryptogrphic key.

\begin{figure}[!htbp]
	\centering
	\includegraphics[width=\textwidth, height=3.3in]{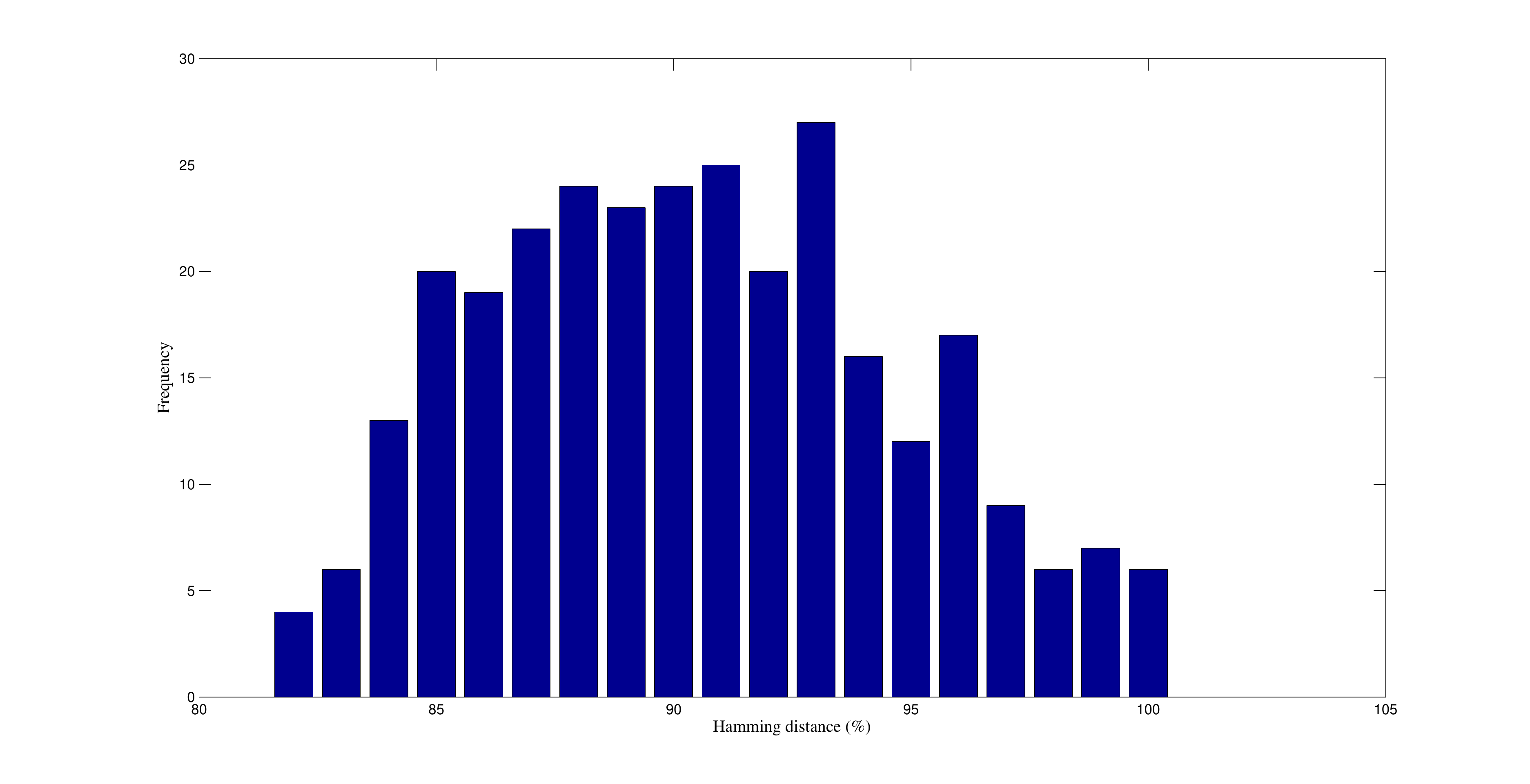}
	\caption{Hamming distances between genuine pair of cryptographic keys}
	\label{fig:f4}
\end{figure}

\subsection{Randomness of cryptographic key for imposter pair of subjects}
To evaluate the randomness in cryptographic keys for imposter pair of users, datasets DB1, DB2, DB3 and DB4 are considered. Each dataset is divided into 50 unique pairs. For each pair, rest of the pairs are considered imposter pairs of fingerprints. This way, a total of $50\times 49$=2450 possible combinations are possible for each genuine pair taking part in communication. Next, we generate the cryptographic keys for each combination of genuine and imposter pair of fingerprints in databases DB1, DB2, DB3 and DB4. Hence, a total of $2450\times 4$=9800 hamming distances have been calculated which are shown in Fig.5 using histogram. It can be observed from the histogram that mean hamming distance is 49.94\% which means that average hamming distance between genuine and imposter keys is 128 bits. Hamming distances are spread between the range of 37.89\% to 61.72\% with a standard deviation of 0.031. Also, it has been observed that, 40\% to 60\% of the bits of genuine keys are different from 99.89\% imposter keys and a small number (0.04\%) of imposter keys have unmatched bits under 40\%. Hence, an imposter cannot get more than 128 matched bits out of any 256-bit cryptographic key.
\begin{figure}[!htbp]
	\centering
	\includegraphics[width=\textwidth, height=3.3in]{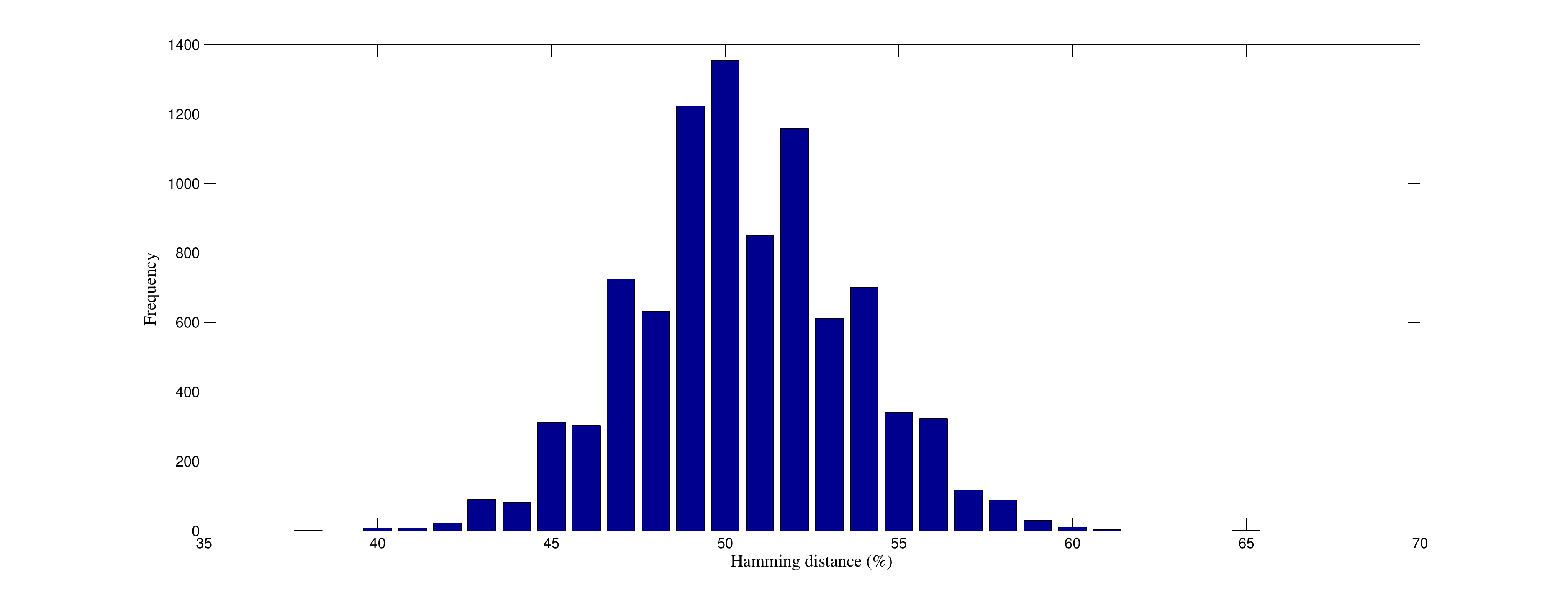}
	\caption{Hamming distances between imposter pair of cryptographic keys keys}
	\label{fig:f5}
\end{figure}

\subsection{Randomness of private key for different transformation key}
To measure the randomness in private keys, we consider all subjects of FVC2002DB1, FVC2002DB2, FVC2002DB3 and FVC2002DB4. Next, private key is evaluated for 30 different randomly generated transformation keys for each subject. Further, we evaluate the Hamming distances between the first private key derived using transformation $T$ and other 30 private keys. This way, a total of 12000 hamming distances are calculated for all subjects in dataset DB1, DB2, DB3 and DB4. Histogram of the Hamming distances is shown in Fig. 6. It can be observed from Fig. 6 that mean hamming distance is 50.03\% which means that average hamming distance between two private keys generated using different transformation keys is 128 bits. Hamming distances are spread between the range of 37.11\% to 64.06\% with a standard deviation of 0.031. For change in transformation keys, 40\% to 60\% bits of private key are different in 99.85\% of cases. Therefore, it can be deduced that at least 128 bits of private key are altered on changing transformation key for a subject.

\begin{figure}[!htbp]
	\centering
	\includegraphics[width=\textwidth, height=3.5in]{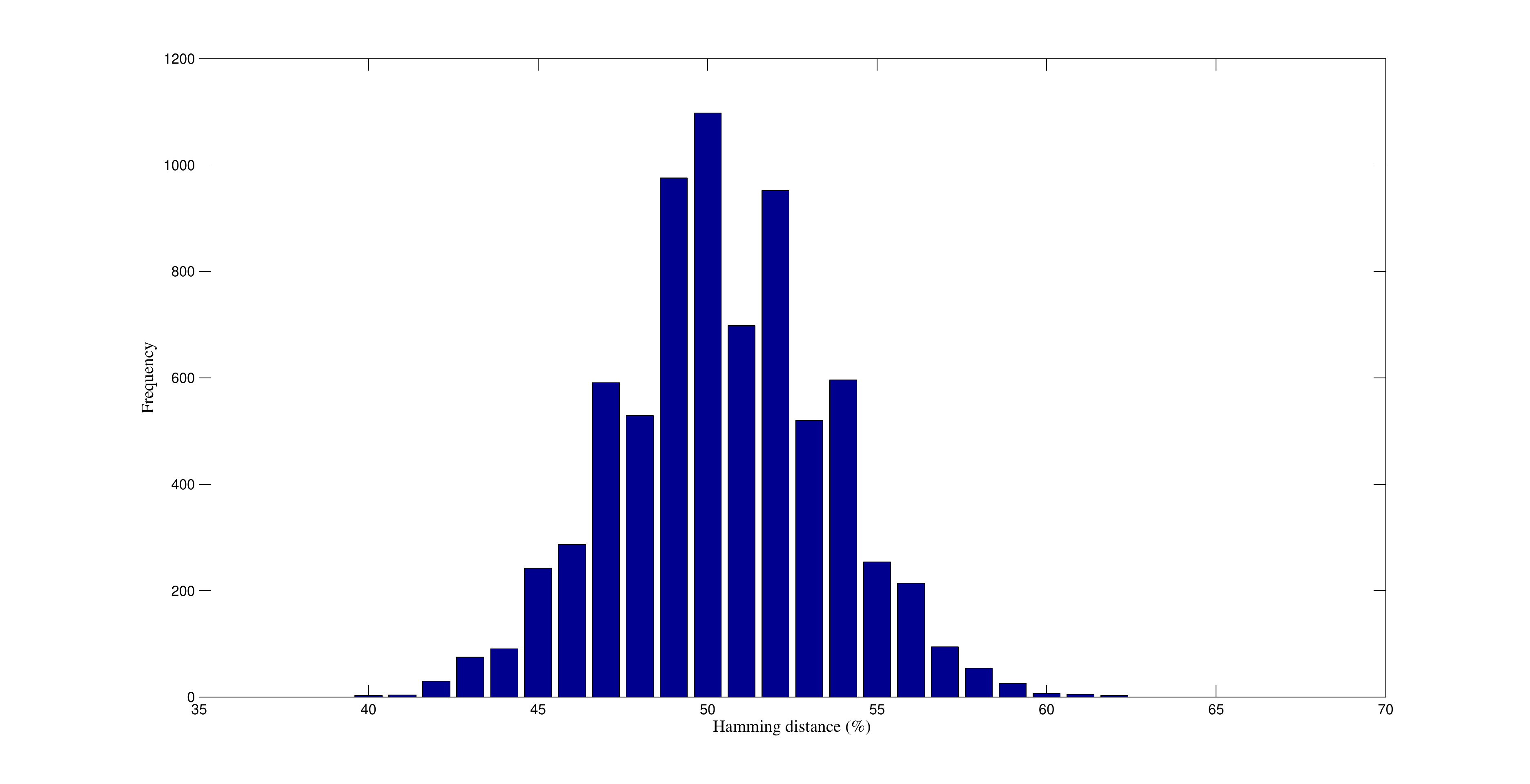}
	\caption{Hamming distances among private keys for different transformation keys}
	\label{fig:f6}
\end{figure}

\subsection{Key size analysis}

In literature, it has been analyzed that an authentication system requires more than $2^{100}$ secret keys according to Alvarez et al. \cite{alva}. In his case, each key must be strong enough and should generate random data to be resilient against an exhaustive attack. In our method, we use 256 bit cryptographic key whose randomness has been tested in Section 3.2-3.4. Hence, $2^{256}$ number of different keys can be derived. The strength of derived cryptographic keys are based upon maximum Lyapunov exponent analysis as stated in Murillo et al. \cite{murillo}.

\subsection{Information entropy analysis}
Information entropy is measured by the randomness present in a cryptographic key, i.e. greater the unpredictability of the key, greater is the entropy. Otherwise, the authentication system is susceptible to an entropy attack since there exists a certain degree of predictability in the key generation. The entropy $H(PUB)$ of a key $PUB$ can be calculated as follows:
\begin{equation}
H(PUB)=\sum_{i=0}^{2^{N}-1}p(PUB_{i})\ log _{2}\left ( \frac{1}{PUB_{i}} \right )
\end{equation}

where, N is the number of bits in the key $PUB$; 2N is all possible bits in key, $p(PUB_{i})$ represents a probability of any bit present in $PUB_{i}$. If there is a key $PUB$ containing $2^{N}$ possible bits, the entropy should be $H(PUB)$=N ideally. The cryptographic key has 256 bits and its maximum entropy is H = 8. This confirms that all bits appear with the same probability. The entropy of the original template is $H$ = 5.14, whereas the entropy of the generated cryptographic key is $H$ = 7.28 for the parameter $n_{p}$=15. Therefore, it is ascertained that the key generation is highly pseudorandom in our method.

\subsection{Performance}
The proposed method has been evaluated using all datasets of (i.e. DB1, DB2, DB3 and DB4) of FVC2002. The datasets cover a wide range of fingerprint images in terms of quality. Among these four datasets, dataset DB3 and DB4 contains the lowest quality images. In this work, minutiae points are extracted using the commercial software VeriFinger SDK \cite{verif}. For each subject, hamming distances between the cryptographic keys generated from genuine pair of fingerprints and hamming distances between the cryptographic keys generated from imposter pair of fingerprints are calculated. For a genuine pair of fingerprints, hamming distances must be minimal while for a pair genuine and imposter fingerprint, hamming distance must be higher. The experimental results are obtained under the practical scenario when transformation key is same for all subjects This scenario represents the stolen token case when an imposter knows the transformation key. We have evaluated the genuine and imposter scores using the same transformation key for each user present in the database. The ROC curves for DB1, DB2, DB3 and DB4 of FVC2002 are shown in Fig. 7. 

\begin{figure}[!htbp]
	\centering
	\includegraphics[width=\textwidth, height=2in]{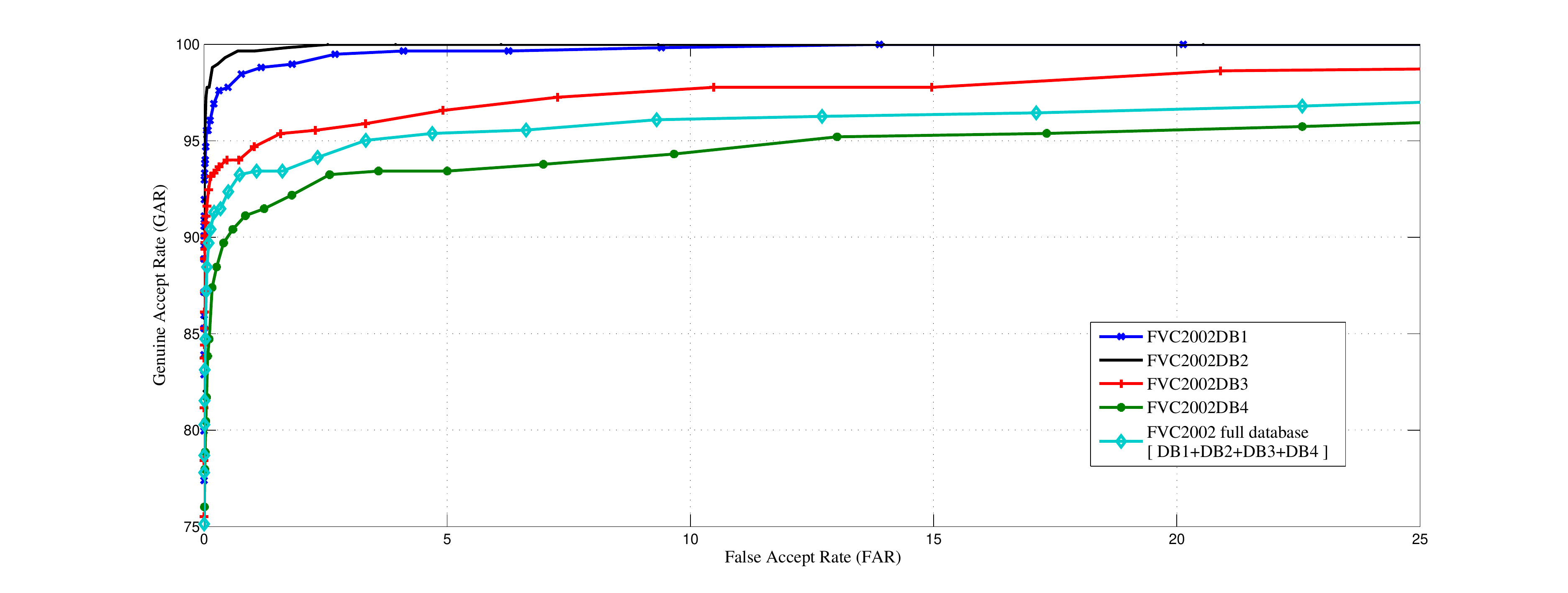}
	\caption{ROC curves for FVC2002 datasets}
	\label{fig:f7}
\end{figure}

\begin{figure}[t]
	\centering
	\includegraphics[width=\textwidth]{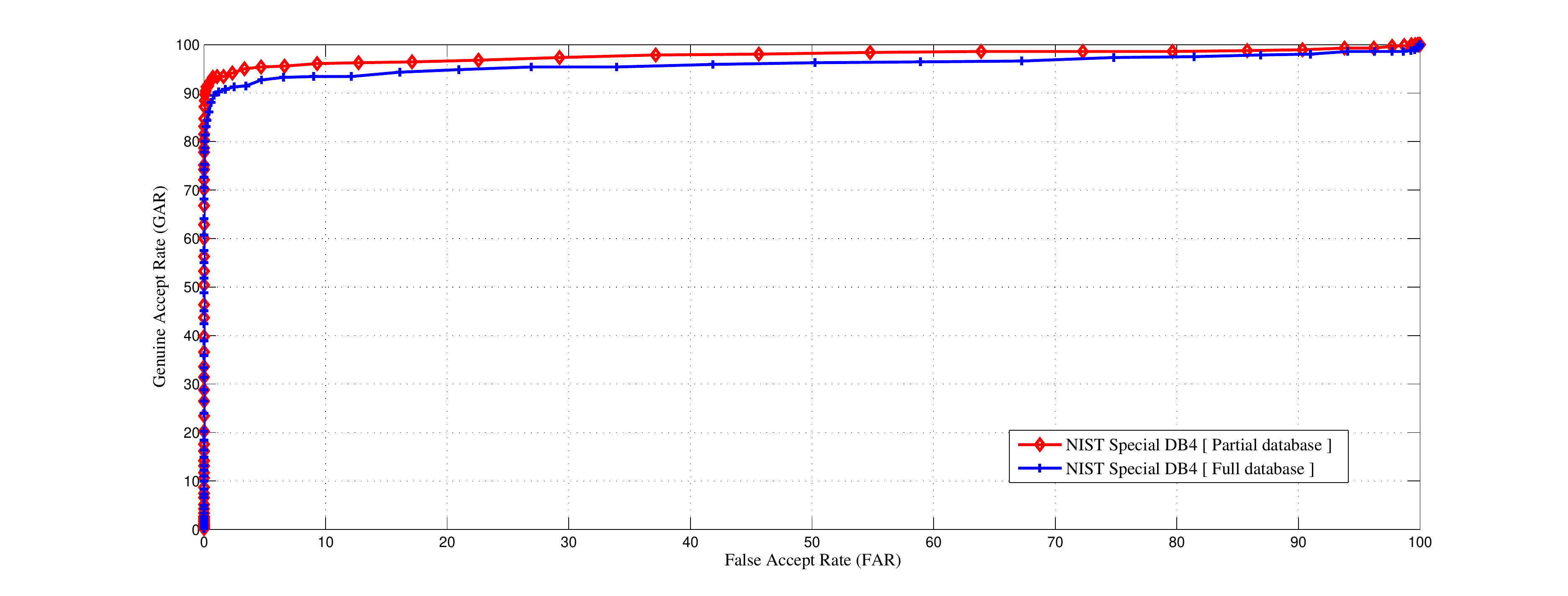}
	\caption{ROC curves for NIST special database 4}
	\label{fig:f8}
\end{figure}

From Fig. 7, it has been observed that we achieve GAR of 98.29\% and 99.03\% for the datasets DB1 and DB2, respectively due to the presence of more number of good quality images. In comparison to DB1 and DB2, relatively less number of minutiae points are extracted from fingerprints in DB3 due to the poor-quality images with spurious and missing minutiae. As a result, we achieve 95.56\% and 86.4\% of GAR for DB3 and DB4, respectively. The lack of reliable minutiae causes the degradation in performance. Further, we also evaluate the performance after combining all the four datasets of FVC2002 resulting into $2800\times 4$=11200 genuine key comparison and $4950\times 4$= 19800 imposter key comparisons. The ROC curve for the whole FVC2002 database is shown in Fig. 7. It has been observed that the GAR of the proposed method is slightly lower than the GAR of the method proposed in \cite{barman2} since error-correction has not been applied for the derived cryptographic keys. Clancy et al. \cite{clancy} and Kanade et al. \cite{session} have not provided any empirical evaluation onto publicly available databases, therefore we have not compared our method with these methods. In addition to the datasets of FVC2002, we also tested our method with NIST special database 4 \cite{nist}. The ROC curve for the NIST special database 4 \cite{nist} is illustrated in Fig. 8. It is worth mentioning that the proposed method achieves a GAR of 96.73\% for partial NIST special database 4, which is better than the reported  95.12\% GAR in \cite{barman3}. However, we obtain high FAR for NIST special database since the error-correction codes have not been applied over cryptographic keys. The performance comparison of the proposed method with some existing cryptosystem design framework is reported in Table 1. We have compared our method with the approaches proposed by Barman et al. \cite{barman2} and Panchal et al. \cite{barman3} due to the scarcity of the research work carried out in this direction and the proposed work achieves the best performance over the current state-of-the-art. From the reported results in Table 1, we can observe that the performance of the proposed method for whole FVC2002 database is slightly lower than \cite{barman2} but it is comparable. 

\begin{table*}[!htbp]
	\centering
	\caption{Comparison with existing crypto-biometric systems \\ GAR ( FAR / FRR )}
	\resizebox{\textwidth}{!}{%
		\begin{tabular}{|c|c|c|c|c|c|c|c|}
			\hline
			\multirow{2}{*}{Methods} & \multicolumn{5}{c|}{FVC2002 database} & \multicolumn{2}{c|}{NIST special database 4} \\ \cline{2-8} 
			& DB1 & DB2 & DB3 & DB4 & DB1+DB2+DB3+DB4 & Partial data & Full data \\ \hline
			Barman et al. \cite{barman2} & - & - & - & - & \begin{tabular}[c]{@{}c@{}}97\\ ( 0.562 / - )\end{tabular} & - & - \\ \hline
			Panchal et al. \cite{barman3} & - & - & - & - & - & \begin{tabular}[c]{@{}c@{}}95.12\\ ( 0 / 4.72 )\end{tabular} & - \\ \hline
			Proposed method & \begin{tabular}[c]{@{}c@{}}98.29\\ (0.2/1.45)\end{tabular} & \begin{tabular}[c]{@{}c@{}}99.03\\ (0.11/0.72)\end{tabular} & \begin{tabular}[c]{@{}c@{}}95.56\\ (1.8/7.73)\end{tabular} & \begin{tabular}[c]{@{}c@{}}86.4\\ (3.08/19.45)\end{tabular} & \begin{tabular}[c]{@{}c@{}}96.49\\ (0.61/2.81)\end{tabular} & \begin{tabular}[c]{@{}c@{}}96.73\\ ( 0.83 / 6.3 )\end{tabular} & \begin{tabular}[c]{@{}c@{}}95.89\\ ( 0.762 / 8.1 )\end{tabular} \\ \hline
	\end{tabular}}
\end{table*}

\section{Security analysis}
In our approach, biometrics of users are neither stored nor shared. Two users can communicate with each other without the worry of storing or sharing their biometric data. In this section, we focus on security of the proposed framework against various possible attacks. Analysis of different attacks show that our approach is robust enough against privacy invasion and security thefts.

\subsection{Security of DH algorithm}
In DH key exchange, it is relatively easy to calculate exponentials modulo a prime while it is very difficult to calculate discrete logarithms. For larger primes, the latter task is considered infeasible \cite{stalling}. DH algorithm requires two parameters q and $\alpha$. For example, prime number q be 353 and its primitive root $\alpha$ be 3. A and B select private keys $X_{A} = {97} \ \text{and} \ X_{B} = {233}$ respectively. Now, the public keys become:
\begin{center}
	$Y_{A} = {3}^{97} \ \text{mod} \ {353} = {40}$
\end{center}
\begin{center}
	$Y_{B} = {3}^{233} \ \text{mod} \ {353} = {248}$
\end{center}
After exchanging public keys, common secret key is: 
\begin{center}
	$K_{r} = (Y_{A})^{X_{B}} \ \text{mod} \ {353} = {248}^{97} \ \text{mod} \ {353} = {160}$
\end{center}
\begin{center}
	$K_{s} = (Y_{B})^{X_{A}} \ \text{mod} \ {353} = {40}^{233} \ \text{mod} \ {353} = {160}$
\end{center}

Assume that, an attacker gets q, $\alpha$, $Y_{A}$ and $Y_{B}$. To evaluate the secret key $K_{r}$ or $K_{s}$, an attacker needs to solve the expression ${3}^{97} \ \text{mod} \ {353} = {40}$ or ${3}^{233} \ \text{mod} \ {353} = {248}$. However, the evaluation becomes impractical for larger primes. Hence, even if an attacker gets access to the public keys, private keys cannot be generated. This ensures that there is no information which the attacker can utilize to reveal the original biometric information of users. 

\subsection{Security of cryptographic key}
In this approach, cryptographic key is generated from biometric of sender and receiver using DH algorithm. This key is valid for only for a session and is destroyed as soon as the session is destroyed. This key is never shared or stored. Therefore, there is no way to get this key by attacking the network. Further, we analyze the security of this system against different possible attacks.

\subsubsection{Network attack}
Assume that, an attacker invades the security of network and takes control over all the information shared over the network. In our method, public keys of sender and receiver are the only information shared over the network before generating a secure session key. Even if an attacker is able to unveil the public keys of sender and receiver, no information can be reverse-engineered using these keys as discussed in section 4.1. 
\subsubsection{Attack on a host}
I this attack, an attacker take control over a user/host in the network and gets all the information available at the user end. In this system, a user stores transformation key, cryptographic session key, public-private key pair and authentication certificate. With all this information, the attacker gets access only for that session. An attacker can log all the encrypted transmissions and can get stored messages of that user, but the attacker still can't decrypt these messages. As cryptographic keys are changed in each session and are not related in any way except that they are generated from original biometric of user, an attacker can access messages of that session only. For decrypting messages of previous communications and to encrypt messages of future conversations, an attacker still needs original biometric of the user which we can assume will be secure with the user. This way, access to cryptographic key of a session gives access to messages of that session only, neither the previous nor the future communications. This property is called perfect forward secrecy which the proposed system achieves.

\subsubsection{Replay attack}
In this attack, a falsified data is injected between the sensor and feature extractor. To avoid this, we utilize the session key between two users. For each session, a different cryptographic key is generated and destroyed after the session gets over. If an attacker eavesdrops a message previously transmitted by genuine users, it would fail since cryptographic key is changed. Even if an attacker eavesdrops one of the public keys shared between two users to launch replay attack, it would not be possible to derive cryptographic keys as it requires user's own private key along with public key. This way, the proposed system found to be secure against replay attacks.

\subsubsection{Man in the middle (MiM) attack}

In MiM attack, the attacker inserts him/herself into a communication between two users, impersonates both users and gains access to information that the two users are trying to send to each other. 
To avoid MiM attack, two users need to verify each other before starting communication. In this method, this verification takes place using certificates provided by trusted certification authority (CA). This verification before communication setup makes sure that a user is communicating with the genuine user at the other end avoiding the MiM. If a MiM eavesdrops two users certificate at the time of verification and sends his certificate to both of them to setup two-way communications simultaneously, he will be verified as himself not the genuine user with which a user wishes to setup the communication. Even for this to happen, a MiM needs to have an authentication certificate provided by the CA which can only be provided to him after verifying his identity. Therefore, an attacker could not be able to get the certificate from CA without identifying himself. Once identified, it would not be possible to launch MiM attack since attacker's identity will be revealed. User's verification from CA prevents the authentication system from MiM attacks in the proposed method.

\section{Conclusion and future scope}
Cryptographic key generation and subsequently its security are the two important issues in traditional cryptography. In this work, we address these issues and provide a novel approach to generate symmetric cryptographic keys using fingerprint biometrics of sender and receiver. We utilize invariant features between pair-minutiae vectors to mitigate the adverse effect over performance due to non-linear deformation at the time of acquisition. To provide revocability to the bit string, random permutation is applied on the bit string based on the user-specific key. Next, the revocable bit strings are fed to DH algorithm along with two predefined parameters to generate public keys of sender and receiver. DH algorithm utilizes user's own private key and other user's public key to generate a session key at both users end. This key is further hashed to generate final cryptographic key. We evaluated our method onto all four datasets i.e.DB1-DB4 of FVC2002 and NIST special database 4. The experimental results demonstrate that the GAR of 96.49\% and 95.89\% for FVC2002 database and NIST special databases, respectively which indicates that our approach performs better than the existing approaches. The proposed method outperforms against different attacks such as network attack, attack on a host, replay attack and MiM attacks.  Thus, this proposed approach provides an effective solution to the need of session-based secure communication setup for transmitting messages over an insecure communication channel. In future, there is a scope of optimizing this approach in terms of performance by applying error-correction codes over the cryptographic keys. In addition, the secure crypto-biometric system design for multimodal biometric systems can be looked into the future.

\section*{Acknowledgments}
We would like to acknowledge Indian Institute of Technology Indore for providing the laboratory support and research facilities to carry out this research. The authors are thankful to SERB (ECR/2017/000027), Deptt. of science \& Technology, Govt. of India for providing financial support to carry out this research work.

\section*{Appendix A}

\subsection*{DH parameters:}
\noindent The following parameters (RFC 3526 : 2048-bit MODP group) were used for implementing DH algorithm.
\\
\noindent 1:] Prime number (for modulo) q: \\

\small
\noindent	FFFFFFFFFFFFFFFFC90FDAA22168C234C4C6628B80DC1CD1  \\
29024E088A67CC74020BBEA63B139B22514A08798E3404DD  \\
EF9519B3CD3A431B302B0A6DF25F14374FE1356D6D51C245 \\
E485B576625E7EC6F44C42E9A637ED6B0BFF5CB6F406B7ED \\
EE386BFB5A899FA5AE9F24117C4B1FE649286651ECE45B3D \\
C2007CB8A163BF0598DA48361C55D39A69163FA8FD24CF5F \\
83655D23DCA3AD961C62F356208552BB9ED529077096966D \\
670C354E4ABC9804F1746C08CA18217C32905E462E36CE3B \\
E39E772C180E86039B2783A2EC07A28FB5C55DF06F4C52C9 \\ 
DE2BCBF6955817183995497CEA956AE515D2261898FA0510 \\
15728E5A8AACAA68FFFFFFFFFFFFFFFF \\
\normalsize
\\
\noindent Primitive root (generator) $\alpha$: 2

\section*{References}

\bibliography{refs}

\begin{thebibliography}{10}
\expandafter\ifx\csname url\endcsname\relax
  \def\url#1{\texttt{#1}}\fi
\expandafter\ifx\csname urlprefix\endcsname\relax\def\urlprefix{URL }\fi
\expandafter\ifx\csname href\endcsname\relax
  \def\href#1#2{#2} \def\path#1{#1}\fi

\bibitem{bioc1}
F.~Hao, R.~Anderson, J.~Daugman, Combining crypto with biometrics effectively,
  IEEE Transactions on Computers 55~(9) (2006) 1081--1088.

\bibitem{bioc}
U.~Uludag, S.~Pankanti, S.~Prabhakar, A.~K. Jain, Biometric cryptosystems:
  issues and challenges, Proceedings of the IEEE 92~(6) (2004) 948--960.

\bibitem{survey}
C.~Rathgeb, A.~Uhl, A survey on biometric cryptosystems and cancelable
  biometrics, EURASIP Journal on Inf. Security 2011~(1) (2011) 3.

\bibitem{monrose}
F.~Monrose, M.~K. Reiter, Q.~Li, S.~Wetzel, Cryptographic key generation from
  voice, in: Proceedings 2001 IEEE Symposium on Security and Privacy, 2001, pp.
  202--213.

\bibitem{hao}
H.~Feng, C.~C. Wah, Private key generation from on‐line handwritten
  signatures, Inf. Management \& Computer Security 10~(4) (2002) 159--164.

\bibitem{irisc}
C.~Rathgeb, A.~Uhl, Context-based biometric key generation for iris, IET
  Computer Vision 5~(6) (2011) 389--397.

\bibitem{chen}
B.~Chen, V.~Chandran, Biometric based cryptographic key generation from faces,
  in: Proceedings of the 9th Biennial Conference of the Australian Pattern
  Recognition Society on Digital Image Computing Techniques and Applications,
  DICTA '07, IEEE Computer Society, Washington, DC, USA, 2007, pp. 394--401.

\bibitem{libor}
L.~Masek, Recognition of human iris patterns for biometric identification,
  Tech. rep., Univ. of Western Australia (2003).

\bibitem{mytec}
C.~Soutar, D.~Roberge, A.~Stoianov, R.~Gilroy, V.~Bhagavatula, Method for
  secure key management using a biometric, {WO} Patent App. PCT/CA1998/000,362
  (Oct.~29 1998).

\bibitem{fucom}
A.~Juels, M.~Wattenberg, A fuzzy commitment scheme, in: Proceedings of the 6th
  ACM Conference on Computer and Communications Security, ACM, New York, NY,
  USA, 1999, pp. 28--36.

\bibitem{vault}
A.~Juels, M.~Sudan, A fuzzy vault scheme, Designs, Codes and Cryptography
  38~(2) (2006) 237--257.

\bibitem{clancy}
T.~C. Clancy, N.~Kiyavash, D.~J. Lin, Secure smartcard based fingerprint
  authentication, in: Proceedings of the ACM SIGMM Workshop on Biometrics
  Methods and Applications, WBMA '03, 2003, pp. 45--52.

\bibitem{threef}
S.~Kanade, D.~Camara, E.~Krichen, D.~Petrovska-Delacretaz, B.~Dorizzi, Three
  factor scheme for biometric-based cryptographic key regeneration using iris,
  in: 2008 Biometrics Symposium, 2008, pp. 59--64.

\bibitem{session}
S.~Kanade, D.~Petrovska-Delacrétaz, B.~Dorizzi, Generating and sharing
  biometrics based session keys for secure cryptographic applications, in:
  Fourth IEEE International Conference on Biometrics: Theory, Applications and
  Systems (BTAS), 2010, pp. 1--7.

\bibitem{barman}
S.~Barman, D.~Samanta, S.~Chattopadhyay, Fingerprint-based crypto-biometric
  system for network security, EURASIP Journal on Information Security 2015~(1)
  (2015) 3.

\bibitem{kana}
S.~G. Kanade, D.~Petrovska-Delacrétaz, B.~Dorizzi, A novel crypto-biometric
  scheme for establishing secure communication sessions between two clients,
  in: 2012 BIOSIG - Proceedings of the International Conference of Biometrics
  Special Interest Group (BIOSIG), 2012, pp. 1--6.

\bibitem{barman2}
S.~Barman, S.~Chattopadhyay, D.~Samanta, G.~Panchal, A novel secure
  key-exchange protocol using biometrics of the sender and receiver, Computers
  \& Electrical Engineering.

\bibitem{barman3}
G.~Panchal, D.~Samanta, S.~Barman, Biometric-based cryptography for digital
  content protection without any key storage, Multimedia Tools and Applications
   1--22.

\bibitem{ks2}
Z.~Jin, A.~B.~J. Teoh, T.~S. Ong, C.~Tee, A revocable fingerprint template for
  security and privacy preserving, KSII Transactions on Internet and
  Information Systems 4 (2010) 1327--1342.

\bibitem{mme}
T.~Kivinen, More modular exponential (modp) diffie-hellman groups for internet
  key exchange (IKE) (2003).

\bibitem{fvc}
D.~Maio, D.~Maltoni, R.~Cappelli, J.~L. Wayman, A.~K. Jain, {FVC}2002: Second
  fingerprint verification competition, in: 16th international conference on
  Pattern recognition, Vol.~3, IEEE, 2002, pp. 811--814.

\bibitem{alva}
G.~ALVAREZ, S.~LI, Some basic cryptographic requirements for chaos-based
  cryptosystems, International Journal of Bifurcation and Chaos 16~(08) (2006)
  2129--2151.

\bibitem{murillo}
M.~A. Murillo-Escobar, F.~Abundiz-P{\'e}rez, C.~Cruz-Hern{\'a}ndez, R.~M.
  L{\'o}pez-Guti{\'e}rrez, A novel symmetric text encryption algorithm based on
  logistic map, 2014.

\bibitem{verif}
S.~Verifinger, Neuro technology (2010).

\bibitem{nist}
C.~I. Watson, C.~Wilson, NIST special database 4 (1992).

\bibitem{stalling}
W.~Stallings, Cryptography and network security: principles and practices,
  Pearson Education,India, 2006.

\end{thebibliography}

\end{document}